\documentstyle[epsfig,aps,twocolumn]{revtex}
\newcommand{\beq}{\begin{equation}}
\newcommand{\eeq}{\end{equation}}
\newcommand{\ba}{\begin{eqnarray}}
\newcommand{\ea}{\end{eqnarray}}
\newcommand{\er}{{\bf r}}
\newcommand{\ka}{{\bf k}}
\begin{document}

\draft
\wideabs{
\title{Quantum degeneracy and interaction effects in spin-polarized 
Fermi-Bose mixtures}

\author{L.\ Vichi$^a$, M.\ Inguscio$^b$, S.\ Stringari$^a$, G.\ M.\ Tino$^c$}
\address{$^a$Dipartimento di Fisica, Universit\`a di Trento
        and Istituto Nazionale per la Fisica della Materia,
        I-38050 Povo,  Italy}
\address{$^b$Dipartimento di Fisica and European Laboratory for
  Nonlinear Spectroscopy (LENS), Universit\`a di Firenze, and Istituto
  Nazionale per la Fisica della Materia, Largo E.\ Fermi 2, I-50125
  Firenze, Italy}
\address{$^c$Dipartimento di Scienze Fisiche, Universit\`a di Napoli
  ``Federico II'' and Istituto Nazionale per la Fisica della Materia,
Complesso Universitario di Monte S. Angelo, via Cintia, I-80126 Napoli, Italy}

\date{\today}
\maketitle

\begin{abstract}
Various features of spin-polarized Fermi gases confined in harmonic
traps are discussed, taking into account possible perspectives of
experimental measurements. The mechanism of the expansion of the gas
is explicitly investigated and compared with the one of an interacting
Bose gas. The role of interactions on the equilibrium and non equilibrium
behaviour of the fermionic component in Fermi-Bose mixtures is
discussed. Special emphasis is given to the case
of potassium isotopes mixtures.
\end{abstract}
}
\narrowtext

\section{Introduction}
\label{sec:intro}

After the achievement of Bose-Einstein condensation (BEC) in
dilute vapors of alkali atoms confined in magnetic traps 
\cite{BEC,BEC2,BEC3}, a challenge for future 
experiments  is the cooling of  samples of fermionic atoms, down to the
degenerate regime.

This perspective has already motivated several theoretical works
(see, for example, \cite{Stoof,Houbiers})
devoted to the study of a possible superfluid phase occurring at very low
temperature, 
analog to the one exhibited by superfluid $^3$He and
superconductors . The behaviour 
of this transition depends crucially on the sign and the size 
of the scattering length and
various predictions have already been made to estimate the value of
the corresponding 
critical temperature. Another peculiar phenomenon exhibited by finite
Fermi gases at low temperature is the occurrence of quantum 
shell effects \cite{Schneider},
bringing  interesting analogies with the physics of atomic nuclei and metal
clusters.
The observation of such phenomena  is seriously limited by the difficulty
of reaching the required regime of very low temperatures. In fact both
superfluidity and shell effects are expected to occur at temperatures much
smaller than the Fermi temperature. For a realistic perspective of experimental
measurements in the immediate future it is consequently useful to explore
quantum degeneracy effects  taking place at
higher temperature. These effects have been already the subject of several
theoretical works
\cite{Fermi,Silvera,Oliva,Butts,Schneider,Molmer,PiTn,Burnett,Jin}.  

A crucial role in the experimental achievement of the low temperature regime
needed to observe quantum degeneracy effects is played by evaporative cooling.
This technique requires frequent collisions in order to ensure
fast thermalization. Scattering between spin-polarized fermions 
in $s$ wave is
inhibited by the Pauli exclusion principle. 
As a consequence one should look for alternative  
scattering processes 
involving different spin states or  atoms belonging to different
species. Fermi-Bose or Fermi-Fermi mixtures are possible candidates.
The resulting scattering processes
will eventually provide  the relevant
thermalization mechanism, allowing for the cooling of the sample 
(sympathetic cooling \cite{Myatt,Timmermans}).

In the following we will mainly discuss the case of Fermi-Bose
mixtures where the effects
of two-body interactions can be significant also on the equilibrium 
properties of the fermionic 
component because of the occurrence of Bose-Einstein condensation which
strongly enhances the value of the central density below  the critical
 temperature,
thereby emphasizing the effects of the mean field interaction. 
In this context we will limit the discussion to the case of spin-polarized 
Fermi gases where the interaction between fermions can be
safely neglected. Notice that in the case of unpolarized Fermi gases also
the interaction between fermions can produce significant effects on
the equilibrium properties if the scattering lenght is
large as happens, for example, in the case of $^6$Li \cite{Stoof,Burnett}.

The aim of the present paper is to discuss some of the key
features exhibited by spin-polarized trapped Fermi gases, with 
special emphasis to the dynamics of the expansion following the
switching off of the trap (Sect.\ \ref{sec1}) and the role of
  interactions in the case of
Fermi-Bose mixtures (Sect.\ \ref{sec2}). We apply the results to the
case of $^{40}$K, the fermionic isotope of potassium, which has been
recently cooled in a magneto-optical trap \cite{Cataliotti}. Potassium
is in fact a good candidate for the experiments proposed here because
of the presence of two bosonic isotopes, $^{39}$K and $^{41}$K 
 in addition to the fermionic
one. The possibility of magnetically trapping these atoms has been
already demonstrated \cite{Prevedelli}. Collisional cross sections for
potassium have been calculated in \cite{Boesten,Cote}. In the
following we assume the values given in \cite{Cote}. The analysis
presented in the present paper can be easily extended to other
fermionic atoms which are presently investigated experimentally as,
for instance, $^6$Li \cite{Abraham}.

\section{THE IDEAL FERMI GAS}
\label{sec1}

The thermodynamic behaviour of ideal trapped Fermi gas has
already been investigated \cite{Silvera,Oliva,Butts,Schneider}. 
In this section we
summarize  the main results and we derive
explicit formulae for the expansion of the fermionic cloud
after turning off the confining potential. These results could be
relevant in view of future measurements on the expanding cloud.

Let us consider $N$ fermions trapped in an axially symmetric
harmonic potential 
\beq \label{extpot}
V( r_{\perp} ,z)=\frac{1}{2} m \left( \omega^2_{\perp}r_{\perp}^2+\omega^2_zz^2
\right)
\eeq
and  ignore  the effects of the two-body interatomic force.
The particle distribution function can be written, in semiclassical
approximation, as
\beq \label{nf}
n(\er,\ka,\beta)=\frac{1}{(2\pi)^3} \frac{1}{\exp(\beta(\frac{\hbar^2k^2}{2m}
+V(r_{\perp},z)-\mu_f))+1},
\eeq
where $\beta=(K_BT)^{-1}$ and $\mu_f$ is the chemical potential, fixed by
the normalization condition
\beq \label{normalizzazione}
N=\int n(\er,\ka,\beta)\,d\er \,\,d\ka=\frac{1}{2  (\hbar \omega)^3}
\int_0^{\infty} \frac{E^2\,dE}{e^{\beta(E-\mu_f)}+1} .
\eeq
In eq.\ (\ref{normalizzazione}) $\omega=(\omega_{\perp}^2 
\omega_z)^{1/3}$ is the geometrical average of
the trapping frequencies.

If more spin states are occupied, then
(\ref{nf}) characterizes the distribution function of each
component separately. It can differ from a component to
another because the number of atoms in each spin state and  the
corresponding trapping potential are in general different. 

At $T=0$ eq.\ (\ref{normalizzazione}) allows one to calculate the Fermi energy
defined by  $E_F=\mu_f(T=0)$. One finds  
\beq \label{fenergy}
E_F = K_BT_F=(6N)^{1/3}\hbar \omega.
\eeq
The Fermi energy (\ref{fenergy}) can be used to define typical length
and momentum
scales \cite{Butts} characterizing the Fermi distribution in coordinate and
momentum space respectively: 
\beq
E_F=\frac 1 2 m\omega_{\perp}^2 R_{\perp}^2 = \frac 1 2 m \omega^2_zZ^2
= \frac{\hbar^2K_F^2}{2m}
\eeq
where $R_{\perp}$ and $Z$ are the radial 
and axial widths of the density
distribution at zero temperature:
\beq \label{nrzero}
n(\er;T=0)=\frac{8}{\pi^2}\frac{N}{R_{\perp}^2Z} \left( 1 - \left({r_{\perp}
      \over R_{\perp}}\right)^2-\left( \frac z Z \right)^2 \right)^{3/2}
\eeq 
while $K_F$ is the width of the corresponding momentum distribution
\beq \label{nkzero}
n(\ka;T=0)= \frac{8}{\pi^2}\frac{N}{K_F^3} \left(1-{k^2 \over
K_F^2} \right)^{3/2} .
\eeq
Eqs. (\ref{nrzero}) and (\ref{nkzero}) hold for positive values of
their arguments.
Eq.\ (\ref{nkzero})
is the analog of the most familiar
momentum distribution $n(k;T=0)=3N/(4\pi K_F^3)\Theta(1-k^2/K_F^2)$ 
characterizing uniform Fermi gases.

\begin{figure}
\epsfig{figure=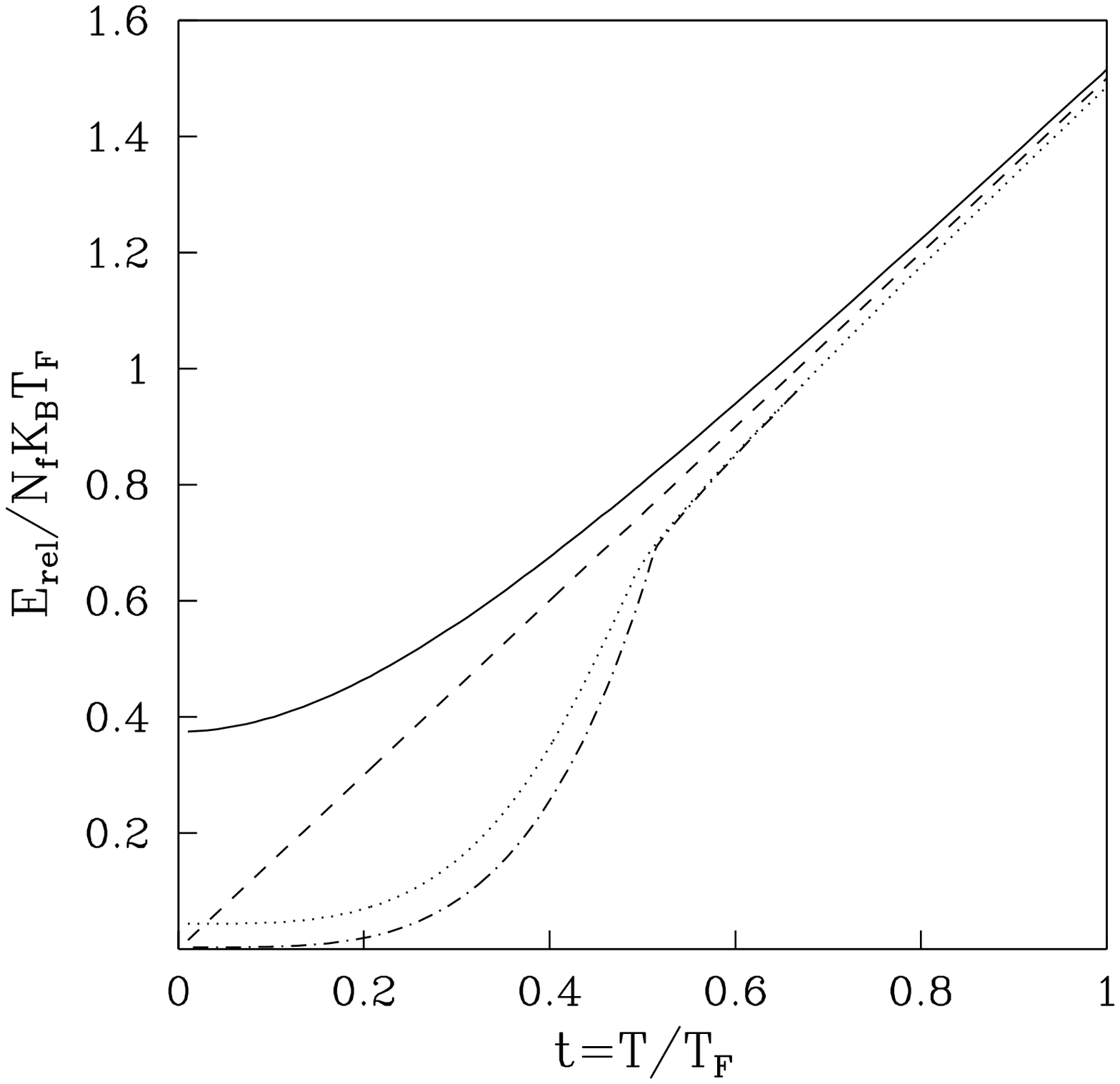,width=0.8 \linewidth}
\caption{}
\label{fig1}
\end{figure}

At finite temperature 
the total energy $E$ of the system  can be written in the form
\beq \label{totalenergy}
E=\frac{1}{2(\hbar \omega)^3}\int_0^{\infty} \frac{E^3\,dE}{e^{\beta(E-\mu_f)}+1},
\eeq
while the release energy, given by the energy of the system after 
switching off the trap
is equal to $E_{rel}=E/2$, because of the equipartition theorem.
At low temperatures one has
$E \simeq\frac{3}{4}NE_F\left(1+ \frac{2\pi}{3}  
\left(\frac{T}{T_F}\right)^2\right)$.
In Fig. \ref{fig1} we 
show the release energy vs $T$ for a gas of
$N$ fermions confined in a harmonic trap (solid line).
In the same figure we also report
the behaviour predicted by the  classical gas ($E_{rel}=3/2NK_BT$,
dashed line) as well as the one
of an ideal Bose gas with the same number of particles $N$,  in the same
confining trap (dot dashed line). For the ideal Bose gas the release
energy is given by
$E_{rel} =1.35  N (K_BT)^4$  
for $T<T_c$ where $T_c= 0.94\hbar \omega
N^{1/3}/K_B$ is the critical temperature for Bose-Einstein
condensation. 
Notice  that, differently from the Fermi case, Bose gases exhibit a
phase transition at $T=T_c$. 
This temperature has the same $N$ dependence
as the Fermi temperature (see eq.\ (\ref{fenergy})) and is fixed by the same
geometrical average of the trapping frequency. The general relation between 
the two
temperatures is given by
\beq
T_F=1.9T_c\left({N_f\over N_b}\right)^{1/3}{\omega_f\over \omega_b}
\eeq
where the suffix $f$ ($b$) refers to fermions (bosons).
The figure clearly shows that in order to observe effects of degeneracy
in the release energy of a Fermi gas 
one should go to temperatures considerably smaller than the Fermi
temperature. For example in order to obtain a deviation of 20 percent from the
classical value one should work at $T \sim 0.3\, T_F$. This represents a major
difference with respect to the thermodynamic behavior of a Bose gas where the
effects of Bose-Einstein condensation in the release energy show up immediately
below the critical temperature as clearly shown in the same figure.

\begin{figure}
\epsfig{figure=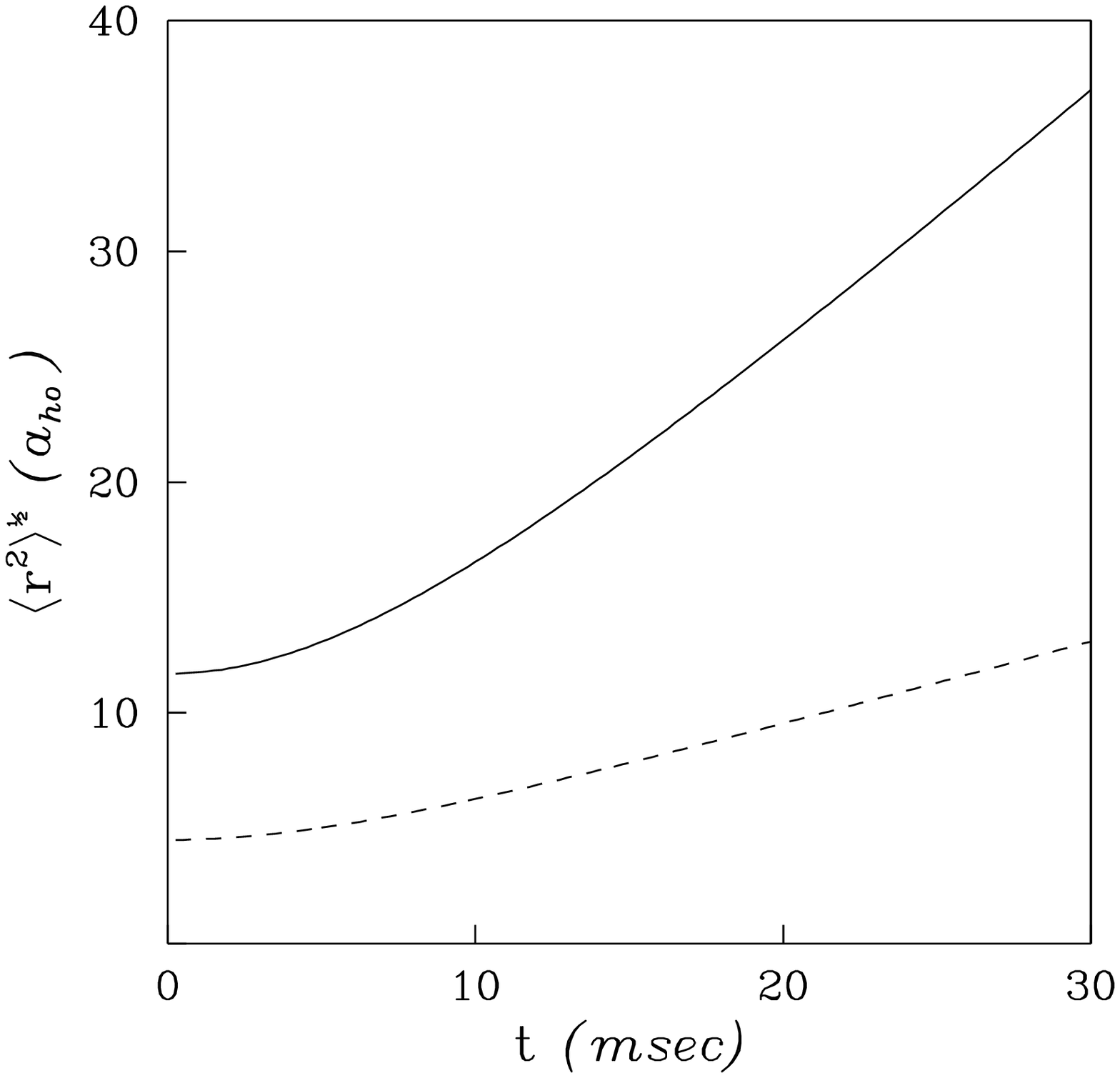,width=0.8 \linewidth}
\caption{}
\label{fig2}
\end{figure}

Let us now consider the expansion of the cloud after turning off the
trapping potential. To obtain the temporal evolution of the
density profile one has to solve the Boltzmann
transport equation. For cold and dilute Fermi gases one can ignore two
body collisions and the distribution function after switching off the
trap will consequently follow the ballistic law
$n(\er,\ka,\beta,t)\equiv n_0(\er-\frac{\ka}{m}t,\ka,\beta)$.
where $n_0$ is the distribution function at $t=0$, given by (\ref{nf}).
One can then easily calculate the time evolution of the
spatial density during the expansion for
which we find the simple analytic result
\ba \label{nrt}
\lefteqn{n(\er,\beta,t)=} && \nonumber \\
&&={6N_f \over R_{\perp}^2Z} \frac{1}{(\pi E_F \beta)^{3/2}}
\frac{1}{(1+\omega_{\perp}^2t^2)(1+\omega_z^2t^2)^{1/2}} f_{3/2}(
\tilde{z})
\ea
where $f_n(z)=(\Gamma(n))^{-1}\int dy\,y^{n-1}/(z^{-1}e^y+1)$ are the
Fermi functions,
$\tilde{z}=\exp(\beta(\mu_f-\tilde V(\er,t)))$ and
\[
\tilde V(\er,t)=\frac 1 2 m\left(\frac{\omega_{\perp}^2r_{\perp}^2}{
1+\omega_{\perp}^2t^2}+
\frac{\omega_z^2 z^2}{1+\omega_z^2t^2}
 \right)
\]
plays  the role of an effective potential fixing, at each
instant, the shape of the distribution function.
The chemical potential has no time dependence, being
fixed by the normalization condition
(\ref{normalizzazione}).\\
From (\ref{nrt}) one can extract the temporal evolution of the mean
square radii of the system:
\beq \label{xyquadmedio}
\langle r_{\perp}^2\rangle = \frac{1}{3N}E_{rel} \frac{2}{m \omega_{\perp}^2}
(1+\omega_{\perp}^2 t^2)
\eeq
\beq \label{zquadmedio}
\langle z^2 \rangle = \frac{1}{3N} E_{rel} \frac{2}{m\omega_z^2}
 (1+\omega_z^2 t^2).
\eeq
which have been written in terms of the release energy  $E_{rel}=E/2$ 
of the gas.
The structure of these equations is  independent of the temperature
  which enters the problem only through the value of the release
energy, fixing the initial value of the widths. Actually the release energy
fixes also the asymptotic behavior of the widths, as clearly shown by
 eqs.\ (\ref{xyquadmedio}-\ref{zquadmedio}).
In Fig. \ref{fig2} we compare the temporal evolution, at $T=0$, 
of the root mean
square radii of the Fermi gas (see eq.\ \ref{xyquadmedio}-\ref{zquadmedio})  
with the one of a gas of interacting bosons. The Fermi gas corresponds
  to $10^6\ ^{40}$K atoms initially confined by a harmonic trap with
  $\omega_f= 100\ Hz$. The Bose gas instead corresponds to $10^6\ ^{39}$K
  atoms interacting with a scattering lenght $a_{bb}=80\ a_0$ \cite{Cote}
  ($a_0$ is the Bohr radius) and initially confined by a magnetic trap
  with $\omega_b=100\ Hz$. The same parameters have been used to
  calculate the release energy of the interacting Bose gas as a
  function of temperature (dotted curve in Fig. \ref{fig1}) in the
  framework of the theory developed in \cite{Giorgini}.
 Fig. \ref{fig2} clearly shows that the Fermi gas expands more quickly
  than the Bose gas due to the significantly higher value of the
  release energy.

Another useful quantity is the aspect ratio of the cloud
\beq \label{aspratio}
R_r(t)=\sqrt{\frac{\langle z^2 \rangle}{\langle x^2 \rangle}}
= \frac{1}{\lambda}\sqrt{\frac{1+\omega_z^2 t^2}{1+\omega_{\perp}^2t^2}}.
\eeq
Eq.\ (\ref{aspratio}) shows that for $t\rightarrow \infty$ the cloud
 becomes spherical in shape even if
it was initially strongly anisotropic. This behaviour is independent of
the temperature and is due to the absence of collisions during the
expansion. Notice that in a Bose gas the situation is quite different,
the asymptotic distribution being anisotropic due to the occurence of
Bose--Einstein condensation, as explicitly pointed out in the first experiments
on BEC \cite{BEC,BEC2}.

\section{Fermi-Bose mixtures}
\label{sec2}

It is important to discuss how the scenario presented in the preceding section
 for the ideal Fermi gas is modified when the system interacts with
a Bose gas confined in the same trap. As discussed in the introduction such
mixed Fermi-Bose gases might become relevant in future experiments for the
achievement of very low temperature regimes via sympathetic cooling.

The study of Fermi-Bose mixtures has been already the subject of theoretical
studies at zero \cite{Molmer} as well as finite  temperature \cite{PiTn}. 
Due to the occurrence
of Bose-Einstein condensation the bosonic component is characterized by a high
density in the central region of the trap where consequently
interaction effects
play an important role. The presence of  the much more dilute Fermi component
 is not expected to influence the bosonic wave function in a
 significant way, so
 that the main effect of interactions between fermions and bosons will result
 in an additional external field acting on the Fermi component. The Fermi gas
 can then, in first approximation, be treated again 
 as an ideal gas trapped by the effective  potential
 \beq \label{veff}
 V_{eff} = V + g_{bf}n_b(r,T)
 \eeq
 where $V$ is the external potential (\ref{extpot}) trapping the
 fermionic species, and the renormalization arises from the
 interaction with bosons. The parameter $g_{bf}=2\pi \hbar^2
 a_{bf}/m_r $
is the Fermi-Bose interaction coupling constant fixed by the relative 
scattering length $a_{bf}$ and by the reduced mass $m_r=m_bm_f/(m_b+m_f)$.
The bosonic density $n_b$ can be calculated
at thermal 
equilibrium using standard procedures developed to describe Bose condensed
gases at finite temperature (see for example \cite{Giorgini}).

The new equilibrium properties of the Fermi gas can be simply understood by
looking at the shape of the potential (\ref{veff}) as explicitly
discussed in \cite{PiTn}. A
first important feature is that the interaction term in (\ref{veff}) 
produces a weaker confinement. This results in an expansion of the Fermi cloud
with respect
to the non interacting case and a consequent decrease of the average
density, as well as of quantum statistical effects.
The effects are more pronounced at low temperature where all the bosons are in
the condensate and one can use the Thomas-Fermi approximation to calculate the
density distribution of the bosonic component. In this case one has
$n_b=15N_b(R_b^2-r^2)/(8\pi R_b^5)$ where 
$R_b=a_{ho}(15N_ba_{bb}/a_{ho})^{1/5}$
 is the classical radius of the Bose gas, $a_{bb}$ is the boson-boson 
 scattering lenght and $a_{ho} =(\hbar/m_b\omega_b)^{1/2}$ is the
 oscillator  length relative to the  bosons. One finds that the
effective potential felt by the fermions is given, for $r<R_b$ by \cite{PiTn}
\beq \label{effpot}
V_{eff} = {m_f\over 2}\omega_f^2\left(1 -{g_{bf} \over g_{bb}} { m_b
    \omega_b^2 \over m_f \omega_f^2 } \right)r^2 +{g_{bf} \over g_{bb}} \mu_b
\eeq
where, for simplicity, we have considered an isotropic trap. For $r>R_b$
the effective potential instead coincides with the bare
potential $V$. Eq.\ (\ref{effpot})
shows that the oscillator constant is reduced by the interaction with bosons.

As a first example we consider
$10^4$ $^{40}$K atoms and $10^6$ $^{39}$K atoms confined in the
same harmonic trap ($\omega_f=\omega_b$).
This corresponds to a
realistic case considering the present status of the experiments
\cite{Cataliotti,Prevedelli}. For the interactions we make the
assumption $g_{bf}\sim 0.5g_{bb}$ \cite{Cote}.

The effect of the interactions is to push the fermionic cloud
``outside'' and to reduce the value of the central density. With the
above choice of parameters we find, at $T=0$, that the central density
is reduced by $\sim 35$ percent, and the average square radius $\langle r^2
\rangle$ increases from $9.5\ a^2_{ho}$ to $11.5 a^2_{ho}$. 
The importance of the interactions on the equilibrium properties of
the Fermi component of the mixture is expected to depend in a crucial
way on the temperature as well as on the relative concentrations 
of the two atomic species. For example at $T=T_F$ one finds a much
smaller effect due to the interactions. In this case the average square radius
increases only by 2.5 \%, even if the central density is reduced by 25
\%. Notice that in the configuration here
considered the Fermi temperature is lower
than the critical temperature
for BEC: $T_F=0.4\,T_c$. This shows that the effects of quantum degeneracy
of the interactions can be observed only at very low temperatures.

Concerning the role of the relative
concentrations we point out that from an
experimental point of view one can access quite a large range of
concentrations. Indeed in \cite{Cataliotti} $10^4$ fermionic potassium
atoms were magneto-optical trapped from a natural abundance
sample. Since samples enriched by two orders of magnitude are
available, $10^6$ $^{40}$K potassium atoms could be easily
loaded. More recently more than $10^8$ bosonic isotopes have been
trapped in a double MOT apparatus \cite{Prevedelli} in the ultra high
vacuum environmental condition required for evaporative cooling. This
method can be easily extended to $^{40}$K. Hence Bose-Fermi mixtures
are expected to be available in the near future with a large variety
of relative abundances.
As a second example we have  considered the case
$N_f=N_b=10^6$. With these values the Fermi temperature is
larger than the BEC temperature ($T_F=1.9T_c$). However the effect of
the interactions are significantly reduced even at $T=0$. For example
the central density has decreased only by 10 \% with respect to the
non interacting case.

\begin{figure}
\epsfig{figure=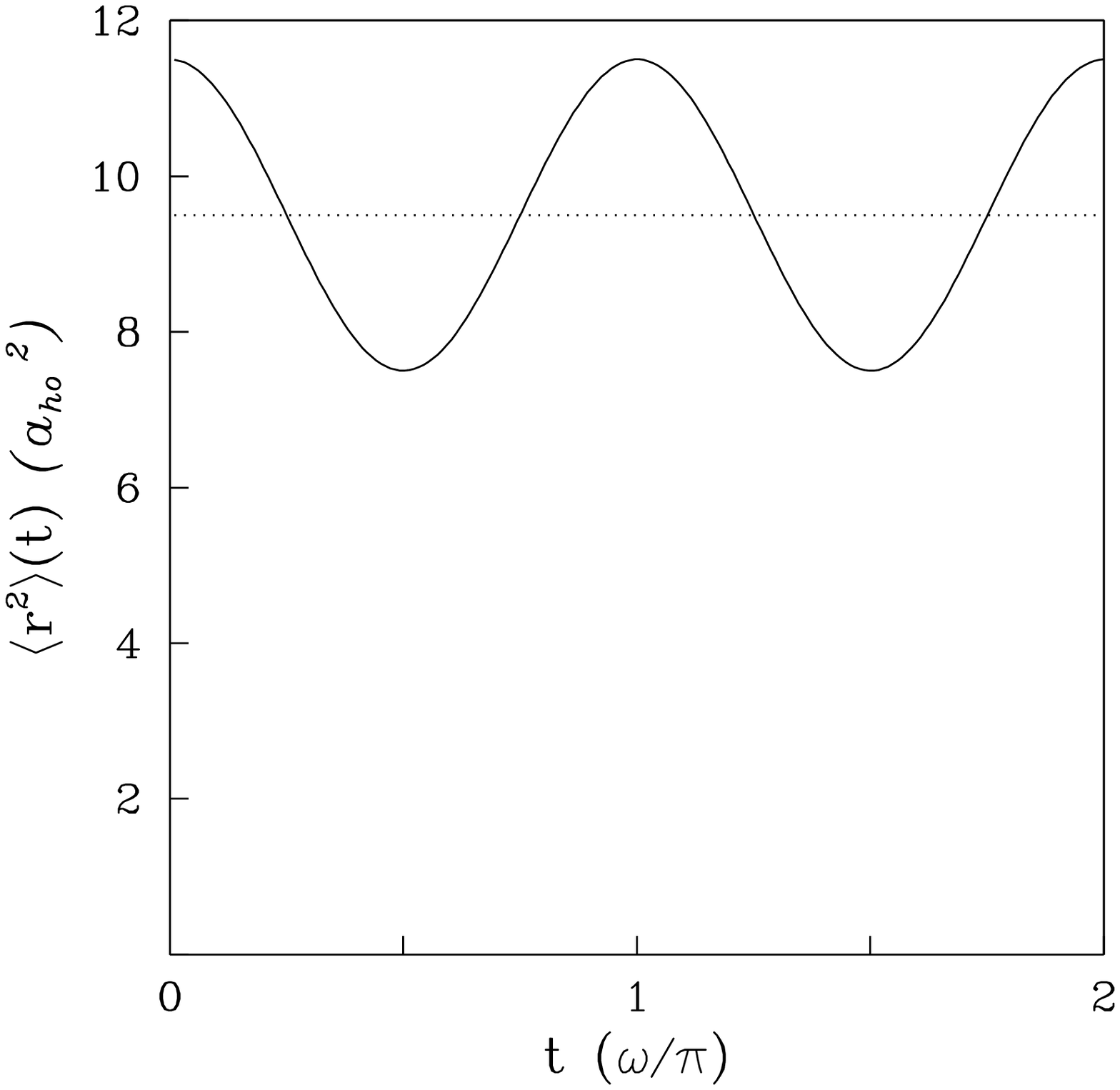,width=0.8 \linewidth}
\caption{}
\label{fig3}
\end{figure}

The above discussion suggests that the main features of
the ideal Fermi gas discussed in section \ref{sec1} are modified only in a semi
quantitative way by the inclusion of the interactions with the bosons.
Nevertheless the effects of the interaction
could be observed experimentally by removing the bosonic
component from the trap at the end of the sympathetic cooling.
This can be realized by optical or rf transitions. 
and corresponds to a
sudden switching off of the interaction term at $t=0$ in the confining
potential (\ref{effpot}). The result is 
an undamped oscillation of the Fermi cloud, shown in Fig. \ref{fig3},
according to the law 
\beq
\langle r^2\rangle (t)= \langle r^2\rangle_0 + \delta 
\cos(2\omega_ft),
\eeq
where $\langle r^2\rangle_0 $ is the equilibrium
mean radius of the fermions in the trap
in absence of interactions, $\delta$ is the difference between the
value of $\langle r^2 \rangle$ calculated at $t=0$ and
$\langle r^2 \rangle_0$. This
oscillation, which is a direct consequence of the Fermi-Bose
interactions, could be observed by imaging the atomic cloud during the
expansion. 

Another consequence
of the interaction between the two components 
is the occurrence of a shift in the frequency of the relative 
motion of the Fermi-Bose clouds, with respect to the predictions of the
non interacting model. This can be easily calculated in the limit where the
number of fermions is much smaller than the one of bosons. In this
case the motion of the fermions is described by the potential (\ref{effpot})
and the center
 of mass of the fermionic cloud oscillates with frequency 
 $\omega= \omega_f\left(1 -{g_{bf} \over g_{bb}} { m_b
    \omega_b^2 \over m_f \omega_f^2 } \right)^{1/2}$.
    With the choice of parameters employed above ($\omega_f
    \sim \omega_b$,
    $m_b\sim m_f$ and $g_{bf} \sim 0.5g_{bb}$) 
the frequency is reduced by about $30$ percent.
 
In conclusion we have shown that the behaviour of a spin-polarized
degenerate Fermi gas, confined in harmonic traps, exhibits several
interesting features, even in the absence of superfluidity. These
features, which are the consequence of Fermi pressure effects, are
directly observable by looking at the expansion of the gas and are
also sensitive to the interactions between fermions and bosons in
atomic mixtures.

We would like to thank E.\ Cornell for useful discussions. This work
was supported by the BEC advanced research project of INFM.

FIGURE CAPTIONS

Fig. 1 Release energy in units of $NK_BT_F$ as a function of temperature (in
units of the Fermi temperature $T_F$). The solid line corresponds to
an ideal Fermi gas 
 trapped in a
harmonic potential. The dashed line is the linear law of a
classical gas, while the
dot dashed and the dotted lines are the release  
energy for a Bose gas with the same number of atoms
confined in the same trap in the non-interacting and in the
interacting case respectively (see the text).

Fig. 2 Temporal evolution in {\it msec} of the root mean square
  radii (in units of $a_{ho}$) of
  an ideal Fermi gas of $^{40}$K (solid line) and of an interacting Bose gas
  $^{39}$K (dashed line) at zero temperature after switching off the
  trap. The two curves refer to a trap
  with $\omega=100\ Hz$
  and same number of particles $N=10^6$. The scattering lenght
  for the $^{39}$K was taken from \cite{Cote}.

Fig. 3 Oscillations at zero temperature
of the mean square radius of $10^4$ $^{40}$K atoms
after the bosons ($10^6$ $^{39}$K atoms) are removed.
Time and lenghts are in units
of $\omega/\pi$ and $a_{ho}$ respectively.

\end{document}